\documentstyle[pra,aps,multicol,graphicx,amsfonts,amstext,amsmath,amssymb,cite,psfrag,color]{revtex}

\newcommand{\CP}{{\mathbb C}{\mathcal P}}
\newcommand{\ddd}{\cdot\!\cdot\!\cdot}
\definecolor{labelkey}{gray}{.1}
\definecolor{refkey}{gray}{.1}
\definecolor{orange}{rgb}{1.000,0.250,0.000}
\definecolor{azure}{rgb}{0.250,0.250,1.000}
\definecolor{dgreen}{rgb}{0.000,0.500,0.000}

\title{Geometric Strategy for the Optimal Quantum Search}
\author{Akimasa  Miyake{\footnote{\tt Email address: miyake@monet.phys.s.u-tokyo.ac.jp}}and Miki Wadati{\footnote{\tt Email address: wadati@phys.s.u-tokyo.ac.jp}}}
\address{Department of Physics, Graduate School of Science, 
         University of Tokyo, \\ 
         Hongo 7-3-1, Bunkyo-ku, Tokyo 113-0033, Japan}
\begin{document}
\draft
\maketitle

\begin{abstract}
We explore quantum search from the geometric viewpoint of a complex 
projective space $\CP$, a space of rays.
First, we show that the optimal quantum search can be geometrically 
identified with the shortest path along the geodesic joining a 
target state, an element of the computational basis, and such an 
initial state as overlaps equally, up to phases, with all the elements 
of the computational basis.
Second, we calculate the entanglement through the algorithm for 
any number of qubits $n$ as the minimum Fubini-Study distance 
to the submanifold formed by separable states in Segre embedding, and
find that entanglement is used almost maximally for large $n$.
The computational time seems to be optimized by the dynamics 
as the geodesic, running across entangled states away from the 
submanifold of separable states, rather than the amount of entanglement
itself.
\end{abstract}
\noindent

%%%%%%%%%%%%%%%%%%%%%%%%%%%%%%%%%%%%%%%%%%%%%%%%%%%%%%%%%%%%%%%%%%%
\begin{multicols}{2}

\renewcommand{\thefootnote}{\arabic{footnote}}
\setcounter{footnote}{0}
%
%  introduction   
%
\section{Introduction}
\label{sec1}
%
%  review
%
Quantum computers would be more powerful than their classical 
counterparts \cite{feynman82,shor94}.
Suppose an oracle function \(f(x)\) with \(x\in\{0,1\}^n\) is given
such that \(f(w)=1\) for an unknown single item \(w\) out of 
\(N(:=2^n)\) and \(f(x)=0\) for \(x\ne w\). Our purpose is to find 
the ``target'' \(w\) with the smallest possible number of the oracle 
evaluations, called the query-complexity. 
As is often the case with computer science, the worst case 
of query-complexity is concerned here. 
If we try with a classical computer, it is readily found that we 
need \(N\) queries in the worst case. On the other hand,    
we can obtain \(w\) with the success probability almost \(1\)
in only \(O(\sqrt{N})\) queries, regardless of \(w\) (i.e. 
for the evaluation in not only the worst case but also the average 
case), by Grover's quantum search algorithm 
\cite{grover97,boyer+98}. 
Furthermore, Zalka \cite{zalka99} proved that Grover's algorithm is 
exactly, and not only asymptotically, optimal for 
query-complexity if quantum computation consists only of unitary 
transformations and the final measurement. \par
Grover's algorithm in \(n\)-qubits (\(2^n\!=\!N\) states) case is 
constructed as follows.
We first introduce an initial ``average'' state \(|a\rangle
:=\frac{1}{\sqrt{N}}\sum_{x=0}^{N\!-\!1}|x\rangle\) where 
\(|x\rangle\;(x=0,\ldots,N\!-\!1)\) forms the orthonormal 
computational basis.
Writing the overlap between the average \(|a\rangle\) and the 
target \(|w\rangle\) by \(\theta\) as
\begin{equation}
\label{theta}
\sin\frac{\theta}{2}:=\langle w|a\rangle=\frac{1}{\sqrt{N}}, 
\end{equation}
we have Grover's algorithm:
\begin{equation}
|a\rangle=\left[\begin{array}{c}
     \cos\frac{\theta}{2}|r\rangle\\ \sin\frac{\theta}{2}|w\rangle 
               \end{array}\right], 
\end{equation}
\begin{align}
\begin{split}
\label{grover-kernel}
G&:=-I_{a}I_{w}
:=-({\bf 1}-2|a\rangle\langle a|)({\bf 1}-2|w\rangle\langle w|) \\
&=\left[\begin{array}{cc}
       \cos\theta&-\sin\theta\\ \sin\theta&\cos\theta      
       \end{array}\right],  
\end{split}\\
\label{grover-psi}
|\psi(k)\rangle&:=G^k |a\rangle
=\left[\begin{array}{c}
       \cos(k\!+\!\frac{1}{2})\theta|r\rangle\\
       \sin(k\!+\!\frac{1}{2})\theta|w\rangle  
       \end{array}\right],
\end{align} 
in the orthonormal basis of \(|r\rangle(:=\frac{1}{\sqrt{N\!-\!1}}
\sum_{x\ne w}|x\rangle)\) and \(|w\rangle\), where
\({\bf 1}\) denotes the \(2n\!\times\!2n\) identity matrix. 
Note that by constructed from alternate inversions for the average 
\(|a\rangle\) and the target \(|w\rangle\) (i.e. \(I_a\) and 
\(I_w\)), the kernel \(G\) in (\ref{grover-kernel}) becomes a real 
\(2\)-dimensional rotation.
We find in (\ref{grover-psi}) that the target \(|w\rangle\) is 
obtained with the success probability \(1\) when 
\((k\!+\!\frac{1}{2})\theta=\pi/2\), i.e. 
\(k\sim\frac{\pi}{4}\sqrt{N}\) by (\ref{theta}) in case of 
\(N\gg 1\). Because one query is used for every \(I_w\) (i.e., 
\(I_{w}|x\rangle=(-1)^{f(x)}|x\rangle\)) of \(G\) in 
(\ref{grover-kernel}), we can thus identify the query-complexity 
with \(k\sim O(\sqrt{N})\). \par 
%
%  motivations
%
Now our motivation is based on the two points below: first,
while it is quite straightforward to verify Grover's algorithm 
\cite{grover97,boyer+98} and Zalka's algebraic proof of its 
optimality \cite{zalka99}, it has yet to be understood from
the geometric aspects why Grover's algorithm works efficiently.
Second, though it is often said that entanglement is useful to 
enhance the quantum information processing, this remains obscure
in theory \cite{braunstein+00} as well as in the Nuclear Magnetic 
Resonance (NMR) experiments \cite{braunstein+99} after, in particular, 
Lloyd's proposition of ``quantum search without entanglement'' 
\cite{lloyd99,meyer00}.
Thus, in this paper, we characterize quantum search from the 
geometric viewpoint, which might shed light on the general 
strategy for constructing efficient quantum algorithms, and discuss 
how entanglement gives quantum computation its power. \par
%
%  organization of the paper
%
The rest of the paper is organized as follows. In Sec.~\ref{sec2},
after we briefly review geometric aspects of quantum mechanics such 
as the complex projective space \(\CP\) and Fubini-Study metric on it, 
we show that Grover's algorithm corresponds to a geodesic of \(\CP\).
In Sec.~\ref{sec3} we discuss entanglement, which can be considered 
as the minimum Fubini-Study distance to the submanifold formed by  
separable states in \(\CP\). Entanglement in Grover's algorithm 
is calculated for the general \(n\)-qubits case, and is found to be 
used almost maximally when \(n\) is large.   
In Sec.~\ref{sec4}, we construct optimal quantum searches, 
including Grover's algorithm, by means of geodesics, and derive 
the geometric necessary and sufficient condition for the optimal 
quantum search. 
Finally Sec.~\ref{sec5} is devoted to conclusions. \par
%
%  preliminary of geometry  
%
\section{Geometric Aspects of Quantum Mechanics}
\label{sec2}
In this section, we first consider, for preliminaries, the pure 
state space of a quantum mechanical system as a complex projective 
space \(\CP\), a space of rays in the associated Hilbert space 
\({\mathcal H}\) \cite{page87,anandan+90,mukunda+93,brody+99,
gibbons92,brylinski00,kus+01}.
Because we discuss the geometric characters of the efficient 
quantum algorithm itself, we can safely restrict to our attentions 
to the pure states.
It implies that we never treat general mixed states (whole states 
given by the density matrix) which appear in some realistic 
situations. After that, we show that Grover's algorithm is the 
horizontal lift of a geodesic in \(\CP\).
\par 
%
%  ray and CP
%
\subsection{Ray and complex projective Hilbert space $\CP$}
Let \(|\psi\rangle\) be a (not necessarily normalized) vector 
in a complex \(N\)-dimensional Hilbert space 
\({\mathcal H}({\mathbb C}^N)\).
The physical state of the quantum system in 
\({\mathcal H}({\mathbb C}^N)\) is given by a ray,
an equivalence class of vectors up to the overall normalization
and phase. So the ray can be interpreted as a line in 
\({\mathbb C}^N\) passing through the origin.
Note that universal quantum computation \cite{deutsch+95,barenco+95}
is defined over rays.
A set of rays forms the complex projective Hilbert space 
\(\CP^{N\!-\!1}\) with the associated projection map \(\Pi\), 
\begin{equation}
\label{proj}
\begin{array}{cccl}
\Pi : & {\mathcal H}({\mathbb C}^N) & \rightarrow & \CP^{N\!-\!1} \\
& |\psi\rangle  & \mapsto & \left\{|\psi'\rangle\mbox{ s.t. }
  |\psi'\rangle=c|\psi\rangle,\;c\in{\mathbb C}\!-\!\{0\}\right\}.
\end{array}
\end{equation}
Suppose \(|\psi\rangle\) is given by \(N\)-tuples 
of complex amplitudes \(z_j\;(j=0,\ldots,N\!-\!1)
\in{\mathbb C}^N\!-\!\{0\}\) by choosing a basis in \({\mathcal H}\).
According to (\ref{proj}), the ray \(\Pi(|\psi\rangle)\) is 
represented as
\begin{equation}
\label{ray-homo}
\Pi(|\psi\rangle)=(z_0',z_1'\ldots,z_{N\!-\!1}'),
\end{equation}
such that \(z_j'=cz_j\) for all \(j\) with \(c\in{\mathbb C}\!-\!\{0\}\).
We find, however, this representation (\ref{ray-homo}), 
called the homogeneous coordinate representation in algebraic geometry, 
is not unique. To obtain a unique one, we also utilize, for any 
nonzero \(z_j\) (say \(z_0\)),  
\begin{equation}
\zeta_l:=\frac{z_l'}{z_0'}=\frac{z_l}{z_0}
\quad (l=1,\ldots,N\!-\!1),
\end{equation}
called the inhomogeneous coordinates. \par 
%
%  Fubini-Study metric
%
\subsection{Fubini-Study metric and geodesics in $\CP$}
Now we introduce Fubini-Study metric, a natural Riemannian metric
in \(\CP^{N\!-\!1}\). Let \(|\psi(s)\rangle\) be a normalized vector 
drawing a curve \({\mathcal C}\) in \({\mathcal H}\) and \(|d\psi(s)\rangle\) 
the tangent vector along \({\mathcal C}\).
Note that the normalization: \(\langle\psi(s)|\psi(s)\rangle\equiv1\)
implies \({\rm Re}\langle\psi(s)|d\psi(s)\rangle=0\).
Under a global gauge transformation: \(|\psi\rangle\mapsto 
e^{i\gamma}|\psi\rangle\) with \(\gamma\in{\mathbb R}\),
the projection, orthogonal to the Hopf fibers, of 
\(|d\psi(s)\rangle\):
\begin{equation}
\label{dpsi-perp}
|d\psi(s)_{\perp}\rangle:=
|d\psi(s)\rangle-\langle\psi(s)|d\psi(s)\rangle|\psi(s)\rangle
\end{equation}
is gauge covariant (i.e. 
\(|d\psi_{\perp}\rangle\mapsto e^{i\gamma}|d\psi_{\perp}\rangle\)).
Since \(\langle d\psi(s)_{\perp}|d\psi(s)_{\perp}\rangle\) is 
gauge invariant, it can be used to define the metric in 
\(\CP^{N\!-\!1}\), called Fubini-Study metric, between two nearby rays 
\(\Pi(|\psi(s)\rangle)\) and \(\Pi(|\psi(s\!+\!ds)\rangle)\) as
\begin{align}
\begin{split}
\label{fs-metric}
\frac{1}{4}ds^2 &:=\langle d\psi(s)_{\perp}|d\psi(s)_{\perp}\rangle\\
&=\langle d\psi(s)|d\psi(s)\rangle 
-\left({\rm Im}\langle\psi(s)|d\psi(s)\rangle\right)^2.
\end{split}
\end{align}
\par 
By variation of the action \(\int_{s_1}^{s_2}ds\) of the line 
element in (\ref{fs-metric}), each extremal gives a geodesic 
\({\mathcal C'}\), which is found to be an arc of the great circle
lying on some submanifold \(\CP^1\) in \(\CP^{N\!-\!1}\) 
\cite{anandan+90,mukunda+93}.
Any lift of the geodesic \({\mathcal C'}\) becomes, 
by definition, a geodesic in \({\mathcal H}\). In particular, 
a horizontal lift of \({\mathcal C'}\), which implies the parallel 
transport: \({\rm Im}\langle\psi(s)|d\psi(s)\rangle=0\),
can be described simply as
\begin{equation}
\label{h-geodesic}
|\psi(s)\rangle=\cos\frac{s}{2}|\psi_1\rangle 
+\sin\frac{s}{2}|\psi_2\rangle, 
\end{equation}
in terms of some orthonormal basis \(|\psi_1\rangle,|\psi_2\rangle\)
in \({\mathcal H}\).
Thus the horizontal geodesic (\ref{h-geodesic}) is just a real 
\(2\)-dimensional rotation on the plane spanned by 
\(|\psi_1\rangle\) and \(|\psi_2\rangle\) in \({\mathcal H}\). 
Furthermore, according to (\ref{h-geodesic}), we can interpret the 
transition probability \(P\) as the distance \(s\;(\in[0,\pi])\)
along the geodesic joining \(|\psi_1\rangle\) and 
\(|\psi(s)\rangle\) \cite{anandan+90,mukunda+93,gibbons92,brody+99}; i.e.  
\begin{equation}
\label{def:fs}
P(|\psi(s)\rangle,|\psi_1\rangle):=|\langle\psi(s)|\psi_1\rangle|^2
=\cos^2\frac{s}{2}.
\end{equation}
We also find the geodesic represents possible superpositions between 
\(|\psi_1\rangle\) and \(|\psi_2\rangle\).
\par
%
%  Grover's algorithm is a geodesic. 
%
\subsection{Grover's algorithm as a geodesic}
If we take \(|\psi_1\rangle=|r\rangle, |\psi_2\rangle=|w\rangle\) 
and \(s=2(k\!+\!\frac{1}{2})\theta\) in (\ref{h-geodesic}), we 
readily find Grover's dynamics (\ref{grover-psi}) satisfies the 
equation of a geodesic in (\ref{h-geodesic}), in addition evolves 
surely along the shorter arc of the geodesic.
This suggests that Grover's dynamics corresponds to the {\em shortest} 
path from the geometric viewpoint. \par
It is significant to note that the original Grover's 
algorithm evolves with discrete \(k\), in other words, it skips
along the geodesic.
The interval of skip becomes shorter as \(N\) becomes larger, 
and almost continuous when \(N\) is sufficiently large.
Here we can regard \(G\) in (\ref{grover-kernel}) as a one-step time
evolution, because we are concerned with the computational 
complexity only in terms of the number of queries called.
This might be called ``coarse-graining,'' where the dynamics driven 
by the detailed physical operations is reduced to the effective
dynamics (i.e. algorithm) of the query-complexity.  \par
%
%  entanglement  
%
\section{Entanglement in Grover's Algorithm}
\label{sec3}
In this section, we explore the geometry of \(\CP\) in more details
to consider relationships between Grover's algorithm and 
entanglement.
As mentioned in Sec.~\ref{sec1}, it is very interesting
whether Grover's algorithm (and general quantum algorithms) takes
advantage of entanglement to compute faster. 
It is shown in \cite{braunstein+00} that Grover's 
algorithm both in the ideal pure state case and in the pseudo-pure 
state in NMR case does generate entanglement during the computation,
by formally tracing out all but one qubit. 
Here we show from the geometric viewpoint that entanglement is used and 
calculate it explicitly. \par
%
%  segre embedding
% 
\subsection{Segre embedding and quadric of separable states}
Some of the mysterious features of quantum mechanics, e.g. 
entanglement and so on, appear when we consider a composite system. 
In the bipartite case, by combining two systems with Hilbert space 
\({\mathcal H}({\mathbb C}^m)\) and \({\mathcal H}({\mathbb C}^{m'})\), 
the combined Hilbert space is taken as the tensor product 
\({\mathcal H}({\mathbb C}^m)\otimes{\mathcal H}({\mathbb C}^{m'})\)
and the associated space of states is \(\CP^{mm'-1}\),
which has a much larger dimension than that of the mere Cartesian 
product \(\CP^{m-1}\!\times\CP^{m'-1}\) (its dimension is only
\(m\!+\!m'\!-\!2\)) of the two individual spaces of states. 
Thus the mysteries seem to lie in the \((m-1)(m'-1)\) relative phases.
Here we consider Segre embedding\cite{gibbons92,brody+99,brylinski00}
in algebraic geometry, which enables products of projective spaces to 
be embedded into a projective space again. Then using the Segre embedding, 
we may characterize entanglement geometrically. \par
We first illustrate the idea in \(2\)-qubits case.
(Segre embedding in the general case is given in 
Appendix~\ref{app:segre}.) 
A state of a qubit is represented by the homogeneous coordinates 
\((z_0,z_1)\in\CP^1\). 
In particular, the spin-up and spin-down basis states \(|0\rangle\) and
\(|1\rangle\) correspond to
\begin{equation}
|0\rangle\leftrightarrow(1,0),\quad |1\rangle\leftrightarrow(0,1),
\end{equation}
respectively (precisely speaking, \(\Pi(|0\rangle)=(1,0)\) and
\(\Pi(|1\rangle)=(0,1)\)).
Then an arbitrary state \((z_0,z_1)\) is a point 
on the complex projective line joining \(\Pi(|0\rangle)\) and 
\(\Pi(|1\rangle)\), interpreted as a superposition of 
\(|0\rangle\) and \(|1\rangle\) with the amplitudes proportional 
to \(z_0,z_1\) respectively as seen in Sec.~\ref{sec2}. \par       
We consider a mapping \(f\) (Segre embedding);
\begin{equation}
\label{map:f}
\begin{array}{rccl}
\lefteqn{f:}&\CP^1\times\CP^1&\rightarrow&\CP^3 \\
&\left((a_0,a_1),(b_0,b_1)\right)&\mapsto&
\left(a_0b_0,a_0b_1,a_1b_0,a_1b_1\right).
\end{array}
\end{equation}
Note that although \((a_0,a_1)\!=\!(\alpha a_0,\alpha a_1),
\;(b_0,b_1)\!=\!(\beta b_0,\beta b_1)\) 
with \( \alpha,\beta\!\in\!{\mathbb C}\!-\!\{0\}\),
the above \(f\) in (\ref{map:f}) map them to the identical 
point in \(\CP^3\), regardless of \(\alpha,\beta\).
Now we discuss the condition for the image of \(f\) :
\(f(\CP^1\!\times\!\CP^1)\) to satisfy in \(\CP^3\).
By writing down the homogeneous coordinates in \(\CP^3\) as 
\((z_0,z_1,z_2,z_3)\), we define a polynomial of degree 2;
\begin{equation}\label{Q}
Q:=z_0z_3-z_1z_2,
\end{equation}
which satisfies \(Q(a_0b_0,a_0b_1,a_1b_0,a_1b_1)=0\).
On the other hand, it is readily checked that arbitrary points on 
\(Q=0\) are included in \(f(\CP^1\!\times\!\CP^1)\). 
Thus we find
\begin{equation}
f(\CP^1\!\times\!\CP^1)=\{(z_0,z_1,z_2,z_3)\;|\;Q=0\}.
\end{equation}
Because we can transform any non-singular quadric 
into the ``normal'' quadric form \(Q=z_0z_3-z_1z_2=0\) by a 
projective transformation \(g\):
\begin{equation}
z_j\stackrel{\textstyle g}{\mapsto}\sum_{l=0}^{N\!-\!1}A_{jl}\:z_l
\quad (j=0,\ldots,N\!-\!1),
\end{equation} 
with an \(N\times N\) matrix \(A:=\{A_{jl}\!\in\!{\mathbb C}\}\) 
s.t. \(\det A\ne0\), we can also identify the non-singular quadric with 
\(\CP^1\!\times\!\CP^1\).   
That is the reason why the algebraic submanifold of separable, or 
no-entangled, states (\(\CP^1\times\CP^1  \)) forms the quadric \(Q=0\) 
in general state space for the 2-qubits system (\(\CP^3\)), and the 
states in \(\CP^3\) off the quadric \(Q=0\) are entangled states. \par
%%%%%%%%%%%%%%%%%%%%%%%%%%%%%%%%%%%%%%%%%%%%%%%%%%%%%%%%%%%%%%%%%%%%%%%%%%%%%
\begin{figure}[bh]
\begin{center}
\begin{psfrags}
\psfrag{w}{\large{\color{magenta}$\zeta$}}
\psfrag{x}{\large $\zeta$}
\psfrag{a}{\large{\color{dgreen}$|a\rangle$}}
\psfrag{b}{\large $|01\rangle$}
\psfrag{c}{\large $|10\rangle$}
\psfrag{d}{\large $|11\rangle$}
\psfrag{e}{\large $|00\rangle$}
\includegraphics[width=5.0cm,clip]{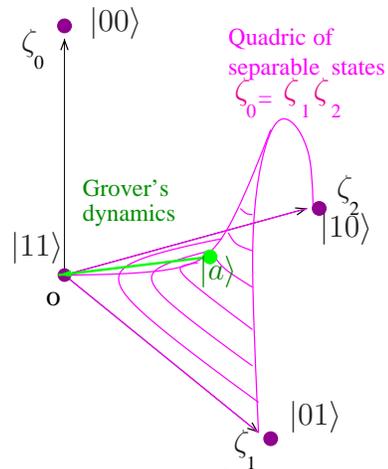}
\caption{The geometry of $\CP^3$ around $|11\rangle$, the assumed target 
$|w\rangle$. Note that because three axes of $\zeta_1,\zeta_2,\mbox{ and }
\zeta_3$ represent complex numbers, this figure is written in the complex 
dimension 3 (real dimension 6). By extracting the real axes of  
$\zeta_1,\zeta_2,\mbox{ and }\zeta_3$, $|11\rangle$ is found to lie on 
a saddle point of the quadric. So is each of the other states of the 
computational basis.}
\label{fig:cp3}
\end{psfrags}
\end{center}
\end{figure}
%%%%%%%%%%%%%%%%%%%%%%%%%%%%%%%%%%%%%%%%%%%%%%%%%%%%%%%%%%%%%%%%%%%%%%%%%%%%%
%
%  usage of entanglement
%
\subsection{Usage of entanglement}
\label{sec3-2}
Let us examine the geometry of \(\CP^3\) in more detail.
We suppose the case where the target state \(|w\rangle\) is 
\(|11\rangle\) without loss of generality.
To draw the behavior around \(|11\rangle
(\leftrightarrow(0,0,0,1))\) as in Fig.~\ref{fig:cp3}, we introduce the
inhomogeneous coordinates: \(\zeta_0=z_0/z_3\), \(\zeta_1=z_1/z_3\),
and \(\zeta_2=z_2/z_3\) because of \(z_3\ne 0\).
Then the quadric \(Q=0\) in (\ref{Q}) is written as 
\begin{equation}
\zeta_0=\zeta_1\zeta_2.
\end{equation}
In particular, all the states orthogonal to \(|11\rangle\), including
\(|00\rangle,|01\rangle,|10\rangle,(|01\rangle+|10\rangle)/\sqrt{2},\) etc., 
are located in points at infinity in Fig.~\ref{fig:cp3}.
The evolution of Grover's algorithm in 
(\ref{grover-psi}) is given by
\begin{align}
\label{psi-2}
\lefteqn{|\psi\rangle} \nonumber\\
&\leftrightarrow
\left(\frac{\cos(k\!+\!\frac{1}{2})\theta}{\sqrt{3}},
\;\frac{\cos(k\!+\!\frac{1}{2})\theta}{\sqrt{3}},\;
\frac{\cos(k\!+\!\frac{1}{2})\theta}{\sqrt{3}},\;
\sin(k\!+\!\frac{1}{2})\theta\right) \nonumber\\
&\leftrightarrow (u,u,u,1),
\end{align} 
or in terms of inhomogeneous coordinates
\begin{equation}
\zeta_0=\zeta_1=\zeta_2=u,
\end{equation}
where \(u:=\cot(k\!+\!\frac{1}{2})\theta/\sqrt{3}\)
changes from \(1\) to \(0\).
We find that Grover's algorithm starts from the average
state \(|a\rangle\;(\zeta_0=\zeta_1=\zeta_2=1)\) on the quadric, 
evolves away from the quadric along a (geodesic) line 
\(0\leq \zeta_0=\zeta_1=\zeta_2 \leq 1\), and finally reaches the 
target \(|w\rangle\) at the origin on it. 
Hence Grover's algorithm uses the entanglement in \(2\)-qubits case. \par
Now we treat the general \(n\)-qubits case.
Recalling (\ref{grover-psi}), we now represent 
\(|\psi(k)\rangle\) by the homogeneous coordinates in 
\(\CP^{N-1}\) (\(N\!=\!2^n\));
\begin{equation}
\label{psi-n}
z_{j\ne w}(k)=\frac{\cos(k\!+\!\frac{1}{2})\theta}{\sqrt{N-1}},\quad
z_w(k)=\sin(k\!+\!\frac{1}{2})\theta.
\end{equation} 
We discuss whether the states of Grover's evolution in 
\(\CP^{2^n-1}\) are included in the algebraic submanifold of the completely 
separable states of \(\CP^1\times\ddd\times\CP^1(=:(\CP^1)^{\times n})\).
As a first step, we consider the condition that (\ref{psi-n}) are 
included in \(\CP^{2^{n\!-\!1}-1}\times\CP^1\). This is just the 
necessary condition for the reduction \(\CP^{2^n-1}\!\rightarrow\!
(\CP^1)^{\times n}\) and, according to (\ref{eq:segre-quadrics}) in 
Appendix~\ref{app:segre} (\(m=2^{n-1}\!-\!1\) and \(m'=1\)), 
is given by  
\begin{equation}
\label{eq:condition}
\frac{\cos(k\!+\!\frac{1}{2})\theta\sin(k\!+\!\frac{1}{2})\theta}
{\sqrt{N-1}}=\frac{\cos^{2}(k\!+\!\frac{1}{2})\theta}{N-1}.
\end{equation}
From (\ref{eq:condition}), we have two cases:
\begin{description}
\item[(i)] If \(\cos(k\!+\!\frac{1}{2})\theta\ne0\), the condition 
(\ref{eq:condition}) becomes
\(\tan(k\!+\!\frac{1}{2})\theta=1/\sqrt{N\!-\!1}.\)
The solutions are given as \((k\!+\!\frac{1}{2})\theta = 
\frac{\theta}{2},\;\frac{\theta}{2}+\pi \pmod{2\pi}\) by use of
(\ref{theta}). 
\item[(ii)] If \(\cos(k\!+\!\frac{1}{2})\theta=0\), it means 
\((k\!+\!\frac{1}{2})\theta=\frac{\pi}{2},\;\frac{3\pi}{2}
\pmod{2\pi}\). These are also the solutions of (\ref{eq:condition}).
\end{description}
The solutions (i) and (ii) are also sufficient, 
i.e. completely separable to \((\CP^1)^{\times n}\), and indeed 
correspond to the average and target state respectively.  
For the states in (\ref{psi-n}) with other \(k\), we cannot 
reduce them  into \((\CP^1)^{\times n}\) and thus they are entangled
states. 
In brief, though the initial (average) state and the 
target state are separable, the intermediate states through which 
the system evolves are entangled. \par 
%
%  calculation of entanglement
%
\subsection{Calculation of entanglement}
\label{sec3-3}
For the Grover's evolution \(|\psi\rangle\), let us calculate 
the amount of entanglement \(E\). Entanglement \(E\) in our pure 
state space is naturally considered the {\em minimum} Fubini-Study 
distance \(s\) to the submanifold formed by completely separable states 
\((\CP^1)^{\times n}\) in \(\CP^{N\!-\!1}\), i.e.
\begin{equation}
\begin{split}
\label{def:entangle}
E(|\psi\rangle)
&:=\min_{\{|\phi\rangle | (\CP^1)^{\times n}\}}
s(|\psi\rangle,|\phi\rangle) \\
&\stackrel{(\ref{def:fs})}{=}
2\arccos\sqrt{\max_{\{|\phi\rangle| (\CP^1)^{\times n}\}}
P(|\psi\rangle,|\phi\rangle)}.
\end{split}  
\end{equation}
Because Fubini-Study distance in (\ref{def:entangle}) can be reduced
from Bures distance with the parallel transport connection in the case of 
pure states\cite{petz+96}, it satisfies the requirements for a good 
measure of entanglement: i.e. (i) zero for any separable state; 
(ii) invariant under local unitary transformations; 
(iii) non increasing expectation value under local operations, 
such as classical communication and subselection,  
as Vedral {\it et al.} suggested in \cite{vedral+97}. \par
It should be remarked that in the case of a {\em bipartite} 
(2-qubits) pure state system, the partial entropy (von Neumann 
entropy of the reduced density matrix associated with one of the 
parties) is widely supposed to be a good measure of entanglement 
\cite{bennett+96}.
However, we apply (\ref{def:entangle}) as the geometric entanglement 
measure, because (i) the partial entropy has no apparent geometric 
meaning in \(\CP\); and (ii) an extension to the multipartite 
(\(n\)-qubits) case is non trivial\cite{footnote0}. 
As a comparison, we calculate, in Appendix~\ref{app:partial},  
the entanglement by the partial entropy in \(2\)-qubits case.
We find our measure of entanglement (\ref{def:entangle}) almost 
corresponds to ``concurrence'' \cite{hill+97} so as to be consistent
with the calculation using the partial entropy. \par
Let us first discuss the \(2\)-qubits case, and then proceed to the
general \(n\)-qubits case. To calculate the entanglement 
of the Grover's state \(|\psi\rangle\) (in (\ref{psi-2})) for 
\(2\)-qubits case, we have to look for the point that gives the 
minimum of \(s\) (or maximum of \(P\)) in (\ref{def:entangle}) on 
the submanifold of the quadric \(Q=0\). Because this point must lie on 
the plane \(\zeta_1=\zeta_2\) (i.e. \(z_1=z_2\)) as seen in 
Fig.~\ref{fig:cp3}, we can parametrize its candidates 
\(|\phi\rangle\) as 
\begin{equation}
\label{coherent-2}
|\phi\rangle\leftrightarrow(v^2,v,v,1),
\end{equation}
with \(v\in{\mathbb C}\) such that \(0\leq |v|\leq1\).
Thus we consider
\begin{align}
\label{maxP}
\lefteqn{\max_{v}P(u,v)} \nonumber\\
&=\max_{v}\frac{[u(v+1)^2 -u+1][u(v^{\ast}+1)^2 -u+1]}
{\left(3u^2 +1\right)\left(|v|^2+1\right)^2} \nonumber\\
&=\max_{(r,\chi)}\frac{(ur^2 + 2ur \cos\chi  +1)^2 
+ 4u(u-1)r^2\sin^2\chi}{(3 u^2 +1)(r^2+1)^2},
\end{align}
where \(v^{\ast}\) denotes the complex conjugate of \(v\) and we use 
\(v=r e^{i\chi}\) with \(0\leq r \leq 1\).
For fixed \(r\), \(P\) in (\ref{maxP}) is the largest
for the phase \(\chi=0\pmod{2\pi}\). This is solely because \(u\) 
in (\ref{psi-2}) is a real number. 
Since \(\frac{\partial }{\partial r}P(u,r,\chi\!=\!0)=0\), we have, 
according to (\ref{def:entangle}),
\begin{equation}
\label{entangle-2}
E(u)=2\arccos\sqrt{\frac{u^2}{3u^2+1}\left(\frac{v_{M}+1}{v_{M}}
\right)^2},
\end{equation}
where \(v_{M}:=[u-1+\sqrt{(u-1)^2+4u^2}]/(2u)\) gives the maximum 
of \(P\) in (\ref{maxP}) with respect to \(r\). \par
\begin{figure}[t]
\begin{center}
\begin{psfrags}
\psfrag{x}{$t\;({\rm rad})$}
\psfrag{E}{{$E(\frac{\cot t}{\sqrt{3}})$}}
\includegraphics[width=7.8cm,clip]{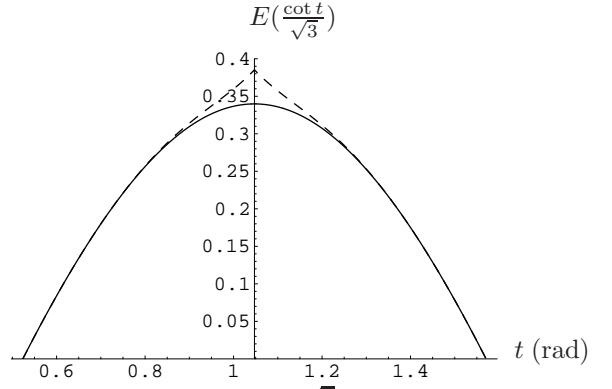}
\caption{Entanglement $E(\cot t/\sqrt{3})$ in (\ref{entangle-2}) during 
Grover's evolution for $2$-qubits case (solid curve) is drawn, compared with
the approximate estimate of the entanglement $E_2(\cot t/\sqrt{3})$ in 
(\ref{entangle-n-app}) (dashed curve). Note that the approximate (dashed) 
curve agrees well with the exact (solid) one except near the half-way state 
\((t=\pi/3)\). }
\label{fig:entangle-2}
\end{psfrags}
\end{center}
\end{figure}

Changing the variable \(u\) into \(t\) by \(u=\cot t/\sqrt{3}\), 
we find the entanglement \(E(\cot t/\sqrt{3})\) changes dynamically 
during the evolution as shown in Fig.~\ref{fig:entangle-2} . 
It takes a value of \(0\) at the initial average
state \(|a\rangle\;(t=\pi/6)\), attains its maximum \(\sim 0.340\) 
at the {\em half-way} state \(u=1/3\;(t=\pi/3)\), and finally goes 
back to \(0\) at the target state \(|w\rangle\;(t=\pi/2)\).  
Note that Grover's algorithm in \(2\)-qubits case uses 
entanglement at most \(\sim 0.340\) though the available maximal 
entanglement is \(\pi/2\).
This implies that, for the half-way state (\(t=\pi/3\)), there is a 
closer state on the quadric than either \(|a\rangle\) or 
\(|w\rangle\) whose distance from the half-way state is 
\(2(\pi/6) >0.34\).
However, as seen in the following, Grover's algorithm comes to use 
the entanglement maximally when the number of the qubit \(n\) 
becomes larger. \par
%%%%%%%%%%%%%%%%%%%%%%%%%%%%%%%%%%%%%%%%%%%%%%%%%%%%%%%%%%%%%%%%%%%%%%%%%%%%%
\begin{figure}[bh]
\begin{center}
\begin{psfrags}
\psfrag{r}{\large $r$}
\psfrag{u}{\large $u$}
\psfrag{h}{\large $u_{\rm half}$}
\psfrag{a}{$n\leq6$}
\psfrag{b}{$n=n_C$}
\psfrag{c}{$n\geq7$}
\includegraphics[width=5.5cm,clip]{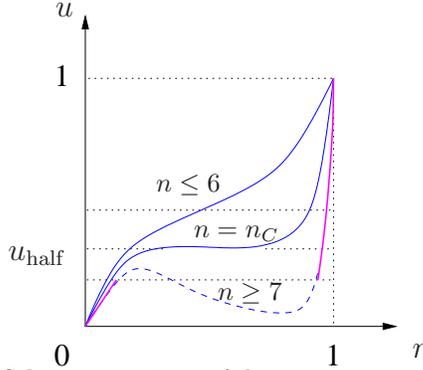}
\caption{Schematic pictures of the extremum condition (\ref{cond:ext1}) for 
\(n\leq6,\) \(n=n_C:=4+2\sqrt{2},\) and \(n\geq7\) are drawn. 
$u_{\rm half}:=\cot[(\pi\!+\!\theta)/4]/\sqrt{N\!-\!1}$ corresponds to the 
half-way state. When \(n\geq7(>n_C)\), the solid parts of the curve give the 
condition for the maximum of $P_n$.}
\label{fig:ext1}
\end{psfrags}  
\end{center}
\end{figure}

%%%%%%%%%%%%%%%%%%%%%%%%%%%%%%%%%%%%%%%%%%%%%%%%%%%%%%%%%%%%%%%%%%%%%%%%%%%%%
Now it is straightforward to calculate the entanglement for the 
general \(n\)-qubits case. According to (\ref{psi-n}), the Grover's 
state \(|\psi\rangle\) in the \(n\)-qubits case is given by 
\((u,\cdots,u,1)\), where 
\(u(:=\cot(k\!+\!\frac{1}{2})\theta/\sqrt{N\!-\!1})\) ranges from 
\(0\) to \(1\). The states \(|\phi\rangle\), candidates closest
to the state \(|\psi\rangle\) on the submanifold \((\CP^1)^{\times n}\), 
are ``coherent'' states (cf. (\ref{coherent-2})) parametrized as
\begin{align}
|\phi\rangle 
&\leftrightarrow\left(v^n,
\underbrace{v^{n-1},\ddd,v^{n-1}}_{n},
\underbrace{v^{n-2},\ddd,v^{n-2}}_{n(n-1)/2},\ddd,
\underbrace{v,\ddd,v}_{n},1\right) \nonumber \\
&\leftrightarrow\underbrace{(v,1)\times\ddd\times(v,1)}_{n}, 
\end{align}
with \(v\in {\mathbb C}\) such that \(0\leq |v|\leq 1\).
Likewise, by use of \(v=r e^{i\chi}\) with \(0\leq r\leq 1\), we have
\end{multicols}
\noindent\rule{0.5\textwidth}{0.4pt}\rule{0.4pt}{\baselineskip}
\begin{equation}
\label{entangle-n}
E_n(u)=2\arccos\sqrt{\max_{v}P_n(u,v)}
=2\arccos\sqrt{\max_{(r,\chi)}\frac{u^2(r^2+2r\cos\chi+1)^n 
+2u(1-u)\sum_{m=0}^{n}{n\choose m}r^m \cos m\chi+(1-u)^2}
{((N-1)u^2+1)(r^2+1)^n}}.
\end{equation}
\hspace*{\fill}\rule[0.4pt]{0.4pt}{\baselineskip}%
\rule[\baselineskip]{0.5\textwidth}{0.4pt}
\begin{multicols}{2}
For a fixed \(r\), the maximum in (\ref{entangle-n}) is attained
at the phase \(\chi=0\pmod{2\pi}\) for the \(n\)-qubits case also.
From \(\frac{\partial}{\partial r}P_n(u,r,\chi\!=\!0)=0\), we have 
an extremum condition;
\begin{equation}
\label{cond:ext1}
u=\frac{r}{(1+r)^{n-1}(1-r)+r}. 
\end{equation}
It is hard to solve analytically the extremum condition 
(\ref{cond:ext1}) for \(r\) so as to seek the solution 
that gives the maximum of \(P_n\). However, we readily find, 
as seen in Fig.~\ref{fig:ext1}, that (\ref{cond:ext1}) increases 
monotonically with \(r\) for \(n\leq6\), on the other hand, it has 
a relative maximum and a relative minimum for \(n\geq7\), 
i.e. for almost all \(n\).
When \(u\) has one-to-one correspondence with \(r\) in 
Fig.~\ref{fig:ext1}, it soon becomes the maximum condition of \(P_n\).
In contrast, when \(u\) has one-to-three correspondence to \(r\), 
the point, among the three, that is included in the solid parts of the curve
in Fig.~\ref{fig:ext1} indeed gives the maximum condition. 
Now it should be noted that because entanglement is symmetric for the half-way
state as in the \(2\)-qubits case, all we need to consider is one half of 
the whole dynamics, e.g. the second half here. 
By reparametrizing \(u\) as \(u=\cot t/\sqrt{N\!-\!1}\), the 
second half, given by \(t\in[(\pi+\theta)/4,\pi/2]\), corresponds to 
\(u\in [\cot[(\pi+\theta)/4]/\sqrt{N\!-\!1},0]\).
Thus when \(N\) is large, the second half is almost 
\(u\in[1/\sqrt{N\!-\!1},0]\) by (\ref{theta}) so that it can be 
treated as the realm of \(u\ll 1\) and \(r\ll 1\).  
Taking first order of \(r\) in (\ref{cond:ext1}), we obtain
an approximate maximum condition for \(r,u\ll 1\): 
\(u\sim r/[1+(n-1)r]\), or
\begin{equation} 
\label{cond:r-max}
r_M:=\frac{u}{1-(n-1)u}.
\end{equation}
Though (\ref{cond:r-max}) becomes a better approximation for  
larger \(N\), it seems to remain valid for small \(N\) because 
even in the \(2\)-qubits (worst approximation) case , the 
deviation from the exact result is limited near the half-way state 
and is small (see Fig.~\ref{fig:entangle-2}). \par
%%%%%%%%%%%%%%%%%%%%%%%%%%%%%%%%%%%%%%%%%%%%%%%%%%%%%%%%%%%%%%%%%%%%%%%%%%%%%%
%%%%%%%%%%%%%%%%%%%%%%%%%%%%%%%%%%%%%%%%%%%%%%%%%%%%%%%%%%%%%%%%%%%%%%%%%%%%%
\begin{figure}[bh]
\begin{center}
\begin{psfrags}
\psfrag{x}{$t\;({\rm rad})$}
\psfrag{E}{{$E_n(\frac{\cot t}{\sqrt{N\!-\!1}})$}}
\includegraphics[width=7.8cm,clip]{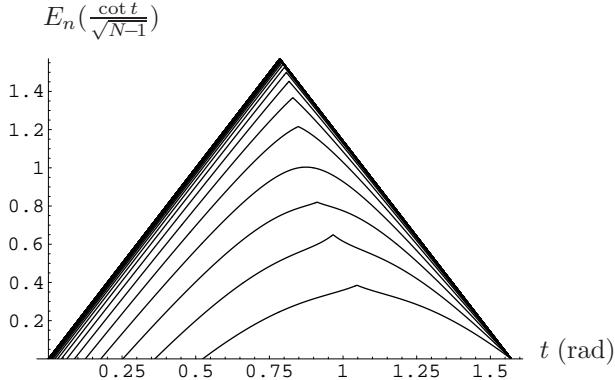}
\caption{Each entanglement $E_n(\cot t/\sqrt{N\!-\!1})$ in 
(\ref{entangle-n-app}) for the $n=2,3,\ddd,50$-qubits case is drawn from the 
bottom to the top. Apparent singularities in the half-way states are just due 
to taking a mirror image of the approximate calculations 
(\ref{entangle-n-app}) for the second half of the dynamics. 
While the true curves for small \(n\) should be smooth near the half-way 
states (cf. Fig.~\ref{fig:entangle-2}), an intrinsic singularity appears as 
the peak of an enveloping triangle $E=-2|t-\frac{\pi}{4}|+\frac{\pi}{2}$ 
when \(n\) goes to infinity.}
\label{fig:entangle-n}
\end{psfrags}
\end{center}
\end{figure}
%%%%%%%%%%%%%%%%%%%%%%%%%%%%%%%%%%%%%%%%%%%%%%%%%%%%%%%%%%%%%%%%%%%%%%%%%%%%%
Substituting \(r_M\) of (\ref{cond:r-max}) and \(\chi=0\) into 
(\ref{entangle-n}), we have the entanglement of the \(n\)-qubits case:
\begin{equation}
\label{entangle-n-app}
E_n(u)\sim 
2\arccos\sqrt{\frac{[u(r_M+1)^n+(1-u)]^2}{[(N-1)u^2+1](r_M^2+1)^n}},
\end{equation}
drawn in Fig.~\ref{fig:entangle-n} with \(u=\cot t/\sqrt{N-1}\).
We find in Fig.~\ref{fig:entangle-n} that  
$E_n(\cot t/\sqrt{N\!-\!1})$ almost converges to an enveloping 
triangle of $E=-2|t-\frac{\pi}{4}|+\frac{\pi}{2}$ at \(n\sim 15\).
This suggests two points: first, entanglement is maximally used
for large \(n\).  Second, the closest separable state during 
Grover's algorithm is either the initial average state 
\(|a\rangle\) or the target state \(|w\rangle\), which implies 
the submanifold of completely separable states is sparse in the large 
\(n\)-qubits state space. \par 
%
%  strategy for optimality  
%
\section{Geometric Construction of Optimal Quantum Search}
\label{sec4}
In Secs.~\ref{sec2} and \ref{sec3}, we found that Grover's algorithm 
is a horizontal lift of a geodesic lying away from the submanifold of 
the separable states in \(\CP\), which can be interpreted as the 
geometric necessary condition for the optimal quantum search. 
In this section, let us consider, on the contrary, whether all 
the geodesics toward the target state \(|w\rangle\) become the 
optimal quantum search. 
That is to say, we discuss the geometric sufficient condition
for the optimal quantum search, from which the bound of the 
computational time is also derived naturally. 
%
%  all geodesics
%
\subsection{Geometric strategy by means of geodesics}
\begin{figure}[b]
\begin{center}
\begin{psfrags}
\psfrag{w}{{\color{magenta}$|w\rangle$}}
\psfrag{r}{{\color{blue}$|r'\rangle$}}
\psfrag{s}{{\color{dgreen}$|y_0\rangle$}}
\psfrag{t}{{\color{dgreen}$|y_1\rangle$}}
\psfrag{u}{{\color{dgreen}$|y_2\rangle$}}
\psfrag{p}{{\color{blue}$|{y_0}_{\perp}\rangle$}}
\psfrag{e}{\large{\color{red}$\frac{\eta}{2}$}}
\includegraphics[width=5.5cm,clip]{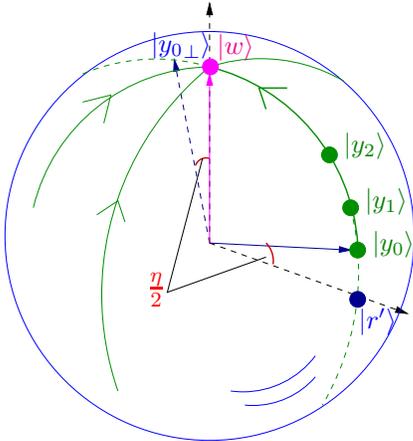}
\caption{Each horizontal geodesic toward the target $|w\rangle$ consists of 
$2$-dimensional successive rotations in the plane including $|w\rangle$ .}
\label{fig:2dim-rot}
\end{psfrags}
\end{center}
\end{figure}

Let us consider a set of all the geodesics through the target state 
\(|w\rangle\) in \(\CP^{N-1}\). As seen in Sec.~\ref{sec2}, 
its horizontal geodesic in \({\mathcal H}\) is just a real 
\(2\)-dimensional rotation on the plane spanned by \(|w\rangle\) and some 
arbitrary state \(|y\rangle\) (denoted for brevity as the \(w\)-\(y\) plane).
We can restrict \(|y\rangle\) such that \(q\;(:=\langle w|y\rangle\in
\mathbb{R})\) ranges from \(0\) to \(1\), by choosing the preferable 
overall phase of \(|y\rangle\) for each ray \(\Pi(|y\rangle)\). 
\(|w\rangle\) and \(|y\rangle\) are said to be ``in phase'' in 
terms of the Pancharatnan connection \cite{mukunda+93}.
By a consequence of an elementary theorem of real Euclidean 
geometry, a \(2\)-dimensional rotation on the \(w\)-\(y\) plane 
is constructed by two successive reflections,
\begin{equation}
\label{kernel-U}
U_{y'}:=-I_{y'}I_{w},
\end{equation}
where \(I_{y'}:={\bf 1}-2|y'\rangle\langle y'|\), 
\(I_{w}:={\bf 1}-2|w\rangle\langle w|\) denote a reflection
for the line orthogonal to \(|y'\rangle\), \(|w\rangle\) in the 
\(w\)-\(y\) plane, respectively.
We take an overall \(-1\) in (\ref{kernel-U}) for convenience,
which simply means that \(-I_{y'}=I_{{y'}_{\perp}}\)
\cite{jozsa99}.
Though in general, \(|y'\rangle\) can be any state on the 
\(w\)-\(y\) plane, we put \(|y'\rangle=|y\rangle\) without loss of
generality because \(|y\rangle\) itself is any state. 
By using \(\eta\in[0,\pi]\) s.t. \(\sin\frac{\eta}{2}:=q
=\langle w|y\rangle\), (\ref{kernel-U}) is represented by,
\begin{equation}
\label{U-eta}
U_{y}(\eta):=-I_{y}I_{w}=\left[\begin{array}{cc}
           \cos\eta & -\sin\eta \\ \sin\eta & \cos\eta  
           \end{array}\right],
\end{equation}
in the basis of \(|r'\rangle(:=(|y\rangle-q|w\rangle)/\sqrt{1-q^2})\),
orthogonal to \(|w\rangle\), and \(|w\rangle\).
Two remarks are in order: first, the angle of the rotation
in (\ref{U-eta}), which corresponds to the speed of a single query, 
is determined just by \(\eta\), (or \(q\)). This means that 
the speed is faster for larger \(\eta\), (or \(q\)). 
Second, the direction of the rotation in (\ref{U-eta}) is 
determined by the order of \(I_w\) and \(I_y\). 
Alternate applications of \(I_w\) and \(I_y\) cause successive 
rotations in the same direction, as can be seen in 
Fig.~\ref{fig:2dim-rot}. \par
Thus the candidate for the algorithm that gives the {\em optimal} 
quantum search toward the target \(|w\rangle\) is constructed in 
terms of the geodesics as
\begin{equation}
\label{Psi-general}
|\Psi(k)\rangle
=U_{y_k}(\eta_k)\cdots U_{y_2}(\eta_2)U_{y_1}(\eta_1)|y_0\rangle,  
\end{equation}
such that \(|y_0\rangle,|y_1\rangle,\ldots,|y_k\rangle\) must lie 
on the {\it same} \(2\)-dimensional plane including \(|w\rangle\) 
with \(\eta_0\leq\eta_1\leq\ddd\leq\eta_k\), where 
\(\sin\frac{\eta_j}{2}:=\langle w|y_j\rangle\).
We find, however, only the case of \(|y_0\rangle=|y_1\rangle=
\ddd=|y_k\rangle\) (i.e. \(\eta_0=\eta_1=\ddd=\eta_k\)) is possible. 
\par
The reason is the following. Suppose the algorithm begins from 
a fixed \(|y_0\rangle\), then
\(|\Psi(1)\rangle=U_{y_1}(\eta_1)|y_0\rangle\) is determined 
by selecting \(|y_1\rangle\) on the \(w\)-\(y_0\) plane.
However, because we never know (only the oracle knows) which is
the target \(|w\rangle\), the only state we are able to utilize 
on the \(w\)-\(y_0\) plane is \(|y_0\rangle\).
So \(|y_1\rangle=|y_0\rangle\).   
Then to get \(|\Psi(2)\rangle\) by the choice of 
\(U_{y_2}(\eta_2)\), it might seem possible to apply
\(|\Psi(1)\rangle\) as well as \(|y_0\rangle\) to \(|y_2\rangle\) 
on the \(w\)-\(y_0\) plane.  
Yet, we must call another oracle as a subroutine to take advantage 
of \(|\Psi(1)\rangle\) in case no measurements are done during the 
computation. This case is just the situation where Zalka
\cite{zalka99} showed the optimality of Grover's algorithm. 
We restrict our attention here to an algorithm including no 
subroutines which require another oracle, because, if needed,
we can always embed no-subroutine algorithm into a certain larger 
algorithm as a subroutine \cite{footnote1}.
Hence all we can do is \(|y_2\rangle=|y_0\rangle\), again. 
In the same way, we finally obtain \(|y_0\rangle=\ddd=|y_k\rangle\)
(i.e. \(\eta_0=\ddd=\eta_k\)) which can be denoted simply by 
\(|y\rangle\) and \(\eta\), respectively. \par
%%%%%%%%%%%%%%%%%%%%%%%%%%%%%%%%%%%%%%%%%%%%%%%%%%%%%%%%%%%%%%%%%%%%%%%%%%%%%%
\begin{figure}[bh]
\begin{center}
\begin{psfrags}
\psfrag{q}{\large{\color{black}$q$}}
\psfrag{t}{\large{\color{black}$T_w$}}
\psfrag{s}{\large{\color{red}$\sim\frac{\pi}{4q}$}}
\psfrag{b}{\large{\color{red}$\sim\frac{\sqrt{2(1-q)}}{\pi}$}}
\includegraphics[width=5.5cm,clip]{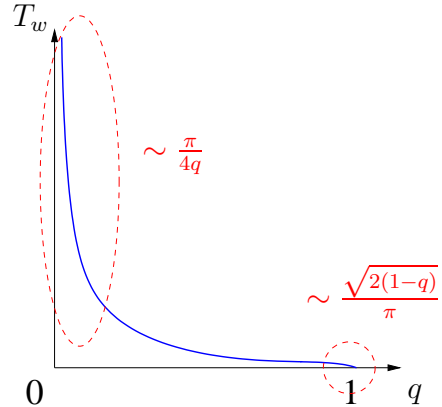}
\caption{The computational time $T_w$ in (\ref{t-w}).}
\label{fig:t-w}
\end{psfrags}
\end{center}
\end{figure}

%%%%%%%%%%%%%%%%%%%%%%%%%%%%%%%%%%%%%%%%%%%%%%%%%%%%%%%%%%%%%%%%%%%%%%%%%%%%%
Accordingly, (\ref{Psi-general}) turns out to be an 
extension of Grover's algorithm where the average state 
\(|a\rangle\) is replaced with the arbitrary state \(|y\rangle\)
\cite{footnote2}. That is, our algorithm is written as
\begin{align}
\begin{split}
\label{geo-alg}
|y\rangle&:=\left[\begin{array}{c}
           \cos\frac{\eta}{2}|r'\rangle \\
           \sin\frac{\eta}{2}|w\rangle
           \end{array}\right], \\
|\Psi(k)\rangle&:={U_y(\eta)}^k|y\rangle=\left[\begin{array}{c}
           \cos(k\!+\!\frac{1}{2})\eta |r'\rangle \\ 
           \sin(k\!+\!\frac{1}{2})\eta |w\rangle 
           \end{array}\right].
\end{split}
\end{align}
The speed of the algorithm (\ref{geo-alg}), considered as the 
traveling (Fubini-Study) distance of a single query along the 
geodesic, is given by 
\begin{equation}
\label{speed}
V(k):=\frac{\Delta s}{\Delta k}
=2\arccos|\langle\Psi (k\!+\!1)|\Psi (k)\rangle|
=2\eta=4\arcsin q,
\end{equation}
where, in the third equality, we use  
\(\langle \Psi(k)|U_y(\eta)|\Psi(k)\rangle=\cos\eta\) by 
(\ref{U-eta}) and (\ref{geo-alg}).
This corresponds to the Anandan-Aharonov relation \cite{anandan+90}:
\(\frac{ds}{dt}=2\Delta H/\hslash\), which means that the speed in
\(\CP\) is determined by the energy uncertainty \(\Delta H\).
Note that \(V(k)\) in (\ref{speed}) is constant, independent of 
\(k\), through the algorithm so that it depends only on \(\eta\) 
(or \(q\)).  
The total traveling distance is naturally thought to be the 
statistical distance \cite{braunstein+94} between the initial
state \(\Pi(|y\rangle)\) and the goal state \(\Pi(|w\rangle)\): 
\begin{equation}
s_w:=\pi-\eta=\pi-2\arcsin q.
\end{equation}
Consequently, we obtain the time required to reach the target
\(|w\rangle\):
\begin{align}
\label{t-w}
T_w:=\frac{s_w}{V}=\frac{\pi-\eta}{2\eta}
=\frac{\pi-2\arcsin q}{4\arcsin q}. 
\end{align}
As seen in Fig.~\ref{fig:t-w}, \(T_w\) is shorter for larger \(q\). 
We also find that \(T_w\sim \pi/(4q)\) for \(q\ll 1\), 
while \(T_w\sim\sqrt{2(1-q)}/\pi\) for \(q\sim 1\). \par
%%%%%%%%%%%%%%%%%%%%%%%%%%%%%%%%%%%%%%%%%%%%%%%%%%%%%%%%%%%%%%%%%%%%%%%%%%%%%
%%%%%%%%%%%%%%%%%%%%%%%%%%%%%%%%%%%%%%%%%%%%%%%%%%%%%%%%%%%%%%%%%%%%%%%%%%%%%%
%
%  bound for optimality
%
\subsection{Bound for the computational time}
We may ask where the bound of the computational time \(T_w\) comes from.
Remember that we want to extract \(|w\rangle\) with probability 1 
in an optimal computational time for the worst case evaluation. 
However, because we do not know {\it a priori} (only the oracle knows)
which is the target \(|w\rangle\), we have to select \(|y\rangle\) 
independently of \(|w\rangle\).
When \(|y\rangle\) is selected as \(|y\rangle=\sum_{x=0}^{N-1} z_{x}|x\rangle
\;(\sum_{x=0}^{N-1}|z_{x}|^2 =1 )\) in the computational basis \(|x\rangle\) 
(\(\ni |w\rangle\)), its smallest overlap \(q_{s}(:=|z_s|)\) gives the 
computational time \(T_{w}\) for the worst case. We find
\begin{equation}
\label{min-t-w}
\frac{\pi - 2\arcsin q_s}{4\arcsin q_s} \geq
\frac{\pi - 2\arcsin (1/\sqrt{N})}{4\arcsin (1/\sqrt{N})}
\sim \frac{\pi}{4}\sqrt{N},  
\end{equation}
where because of \(q_s \leq 1/\sqrt{N}\), the equality in (\ref{min-t-w}) 
is attained for the \(|y\rangle\) such that \(q_s=1/\sqrt{N}\), i.e., 
all \(q_x:=|z_x|=1/\sqrt{N}\). This implies that the ``mixedness'' of 
the searching state space (in part) bounds the efficiency of quantum search 
as Bose {\em et al.} \cite{bose+00} mentioned.
Thus, regardless of which is the target, the optimal computational 
time is \(\frac{\pi}{4}\sqrt{N}\).
This result of course coincides with Grover's result \cite{grover97}, 
first proved optimal by Zalka \cite{zalka99}.\par
It should be commented that there remains room for 
relative phases in \(|y\rangle\). \(|y\rangle=|a\rangle\) in the
original Grover's algorithm is only a choice. In general, any 
element of the Fourier basis, 
\begin{equation}
|p\rangle:=\frac{1}{\sqrt{N}}\sum_{x=0}^{N-1}
e^{\frac{2\pi i}{N}px}|x\rangle
\quad (p=0,1,\ldots,N\!-\!1),
\end{equation}
can be taken as a \(|y\rangle\) among the completely separable states. 
This implies that the quantum search takes advantage of 
dual bases, \(|x\rangle\) and \(|p\rangle\), to run in the 
optimal computational time because its kernel takes the form 
\(U=-I_{p}I_{x}\) by (\ref{U-eta}).  
%
%  conclusions
%
\section{Conclusions}
\label{sec5}
In this paper, we have shown two geometric characteristics of the quantum
search: one is related to the geodesic in \(\CP\), the other
related to entanglement.
First, the geometric necessary and sufficient condition for the 
optimal quantum search is given by the horizontal geodesic joining 
the target \(|w\rangle\) and a preferable selected initial state 
\(|y\rangle\) such that it overlaps equally, up to relative phases, 
with all the elements of the computational basis 
\(|x\rangle \;(\ni |w\rangle)\). 
Second, Grover's quantum search uses entanglement for an arbitrary 
number of the qubits \(n\), in particular almost maximally for large \(n\). 
However, there seems to be no direct relationship between the amount of 
entanglement (how far the dynamics is away from the submanifold of 
separable states) and the optimal, i.e. shortest, computational 
time. This is because (i) the amount of entanglement is different 
for each \(n\) though Grover's algorithm is exactly optimal 
regardless of \(n\) \cite{zalka99}; (ii) the computational time is 
rather determined by the overlap \(q=\langle w|y\rangle\) as seen in 
Sec.~\ref{sec4}.
It is significant that the algorithm consists of
the shortest path by means of the geodesic; as a result it runs 
across entangled states away from the submanifold of the separable states. \par
It is readily found that the multiple target case \cite{boyer+98} is
also characterized in completely the same manner. Moreover, our 
geometric strategy would be useful to construct other efficient 
quantum algorithms, as some efficient classical algorithms are 
widely known to be geodesics in their parameter spaces. 
Exploring the geometric viewpoint also seems appealing toward the realization
of quantum computers. 
For instance, 
(i) the holonomic approach to quantum computation\cite{pachos+01}, 
where loops by horizontal lifts of the path in \(\CP\) construct the 
logic gates to compute quantum algorithms,
is supposed to have built-in fault-tolerant features against local 
perturbations;  
and (ii) time optimal pulse sequences in NMR 
quantum computing\cite{khaneja+01}, given by geodesics on certain coset
spaces, would minimize the effect of relaxation and optimize the sensitivity 
of the experiments. \par
%
%  acknowledgments
%
\section*{ACKNOWLEDGMENT}
One of the authors (A.M.) would like to thank I. Tsutsui for kind
interest in this work.
%
%  appendices 
%
\appendix
%
%  Segre embedding
%
\section{Segre Embedding in the General Case}
\label{app:segre}
It is straightforward to extend the mapping \(f\) of the Segre 
embedding into the general case:
\begin{equation}
\begin{array}{rccl}
\lefteqn{f:}&\CP^{m}\times\CP^{m'}&\rightarrow&\CP^{(m+1)(m'+1)-1}\\
&\left((a_0,\ddd,a_m),(b_0,\ddd,b_{m'})\right)&
\mapsto&\left(a_0b_0,\ddd,a_0b_{m'},a_1b_0, \right.\\
&&&\left. \ddd,a_mb_0,\ddd,a_mb_{m'}\right).
\end{array}
\end{equation}
We find, in the same way as in the text, that the algebraic submanifold 
given by the image of \(f\) is the zero locus of all the homogeneous 
polynomials of degree 2:
\begin{equation}
Q_{(i,j),(k,l)}
:=z_{(m'+1)i+k}z_{(m'+1)j+l}-z_{(m'+1)i+l}z_{(m'+1)j+k},
\end{equation}
where \(0\leq i< j\leq m,\;0\leq k< l\leq m'.\)
Hence we have 
\begin{equation}
\label{eq:segre-quadrics}
f(\CP^m\times\CP^{m'})=\{Q_{(i,j),(k,l)}=0\},
\end{equation}
where a set of quadratic constraints \(\{Q_{(i,j),(k,l)}=0\}\) 
consists of \(m(m+1)m'(m'+1)/4\) simultaneous equations. 
%
%  entanglement by partial entropy
%
\section{Calculation of Entanglement by Partial Entropy in 2-qubits
case}
\label{app:partial}
In this appendix, the entanglement of the 2-qubits case is calculated in
terms of the partial entropy, so as to be compared with the results 
in Sec.~\ref{sec3}.
For the Grover's state \(|\psi\rangle=(u|00\rangle\!+\!u|01\rangle\!
+\!u|10\rangle\!+\!|11\rangle)/\sqrt{3u^2\!+\!1}\) in (\ref{psi-2}), 
we obtain a reduced density matrix by tracing out e.g. the second 
qubit;
\begin{equation}
\label{red-density}
\rho_{\rm red}
:=\mbox{tr}_{\rm 2nd}\left(|\psi\rangle\langle\psi|\right)
=\frac{1}{3u^2+1}\left(\begin{array}{cc}
2u^2&u(u+1)\\u(u+1)&u^2+1
\end{array}\right).
\end{equation}
We calculate entanglement as the partial entropy of the 
reduced density matrix (\ref{red-density}):
\begin{equation}
\label{entangle-p} 
e(u):=-\mbox{tr}\left(\rho_{\rm red}\log\rho_{\rm red}\right)
=-\lambda_{+}\log\lambda_{+}-\lambda_{-}\log\lambda_{-},
\end{equation}
where \(\lambda_{\pm}\), the eigenvalues of the reduced density 
matrix \(\rho_{\rm red}\) in (\ref{red-density}), are given by
\begin{align} 
\lambda_{\pm}&:=\frac{1\pm\sqrt{1-C(u)^2}}{2},
\label{lambda} \\
C(u)&:=2\left|\mbox{det}\frac{1}{\sqrt{3u^2+1}}
\left(\begin{array}{cc} u&u\\u&1\end{array}\right)\right|
=\frac{2u(1-u)}{3u^2+1}.
\label{c}
\end{align}
As \(C(u)\) in (\ref{c}), called ``concurrence''\cite{hill+97}, 
goes from \(0\) to \(1\), the entanglement \(e(u)\) in 
(\ref{entangle-p}) increases monotonically from 0 to 1. 
So \(C(u)\) as well as \(e(u)\) can be regarded as a measure of 
entanglement. \par
\begin{figure}[b]
\begin{center}
\begin{psfrags}
\psfrag{t}{$t\;({\rm rad})$}
\psfrag{e}{$e(\frac{\cot t}{\sqrt{3}}),C(\frac{\cot t}{\sqrt{3}}),
E(\frac{\cot t}{\sqrt{3}})$}
\includegraphics[width=7.8cm,clip]{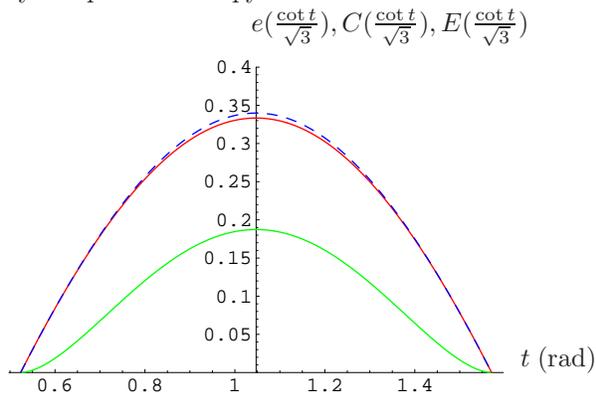}
\caption{Three entanglement measures: (i) the partial entropy 
$e(\cot t/\sqrt{3})$ in (\ref{entangle-p}) (the lower green solid curve); 
(ii) the concurrence $C(\cot t/\sqrt{3})$ in (\ref{c}) (the upper red solid 
curve); and (iii) the minimum Fubini-Study distance to separable states 
$E(\cot t/\sqrt{3})$ in (\ref{entangle-2}) (the upper blue dashed curve).}
\label{fig:pentropy-concur}
\end{psfrags}
\end{center}
\end{figure}

The partial entropy \(e(\cot t/\sqrt{3})\) as well as the concurrence 
\(C(\cot t/\sqrt{3})\), compared with our entanglement 
\(E(\cot t/\sqrt{3})\) in (\ref{entangle-2}), are drawn in 
Fig.~\ref{fig:pentropy-concur}.
We note two points in Fig.~\ref{fig:pentropy-concur}: First, all
the three entanglement measures are convex upward and are maximized 
at the half-way state \(t=\pi/3\;(u=1/3)\). 
Second, surprisingly enough, our entanglement measure \(E\) almost 
coincides with the concurrence \(C\), characterized also as 
\(C=2|z_0 z_3 -z_1 z_2|=2|Q|\) (in (\ref{Q})) with normalized homogeneous 
coordinates \(z_j\).
It should be noted, however, that their normalizations are different, 
i.e. \(E\) is normalized to \(\pi/2\), while \(C\) is normalized to \(1\). 
In summary, our calculation of entanglement by 
(\ref{def:entangle}) is consistent with that by the partial entropy. \par
%%%%%%%%%%%%%%%%%%%%%%%%%%%%%%%%%%%%%%%%%%%%%%%%%%%%%%%%%%%%%%%%%%%%%%%%%%%%%%
%%%%%%%%%%%%%%%%%%%%%%%%%%%%%%%%%%%%%%%%%%%%%%%%%%%%%%%%%%%%%%%%%%%%%%%%%%%%%%
%
%  references  
%

\end{multicols}

\begin{thebibliography}{~~}
\bibitem{footnote1}
The dynamical improvement of \(|y_j\rangle \) in which \(\eta_j\) 
became larger would make the algorithm faster if another oracle were
not required.
In fact, because using other oracles in a subroutine can be found 
to result in an equivalent algorithm to the original Grover's one 
for the computational time, it does not lead to a speed-up.
\end{thebibliography}
\end{document}